# Engineering Kondo state in two-dimensional semiconducting phosphorene

Rohit Babar[1] and Mukul Kabir[1, 2, *]

[1]*Department of Physics, Indian Institute of Science Education and Research, Pune 411008, India*
[2]*Centre for Energy Science, Indian Institute of Science Education and Research, Pune 411008, India*

Correlated interaction between dilute localized impurity electrons with the itinerant host conduction electrons in metals gives rise to the conventional many-body Kondo effect below sufficiently low temperature. In sharp contrast to these conventional Kondo systems, we report an intrinsic, robust and high-temperature Kondo state in two-dimensional semiconducting phosphorene. While absorbed at a thermodynamically stable lattice defect, Cr impurity triggers an electronic phase transition in phosphorene to provide conduction electrons, which strongly interact with the localized moment generated at the Cr site. These manifests into intrinsic Kondo state, where the impurity moment is quenched at multi-stage and at temperatures in the 40–200 K range. Further, along with a much smaller extension of Kondo cloud, the predicted Kondo state is shown to be robust under uniaxial strain and layer thickness, which greatly simplifies its future experimental realization. We predict the present study will open up new avenues in Kondo physics and trigger further theoretical and experimental studies.

## I. INTRODUCTION

Kondo effect observed in non-magnetic metals with dilute magnetic impurities was an early manifestation of strongly correlated electron system, where the electrical resistance and magnetic properties are severely altered at low temperature. [1–3] Rather than saturation, the resistivity increases as the temperature is lowered below a critical temperature known as Kondo temperature $T_K$. Further, the impurity magnetic moment is screened below $T_K$ by the formation of a many-body singlet state with the itinerant conduction electrons of the metal host. This serves as scattering centre for conduction electrons resulting in anomalous temperature dependence in resistance.

Beyond the traditional metal-impurity systems, a wide range of artificial Kondo systems have been realized with the advent in nano-fabrication. Composite systems including single molecules, [4–6] carbon nanotubes [7, 8] and quantum dots [9, 10] attached with metallic electrodes/tunnel-junctions are found to exhibit Kondo resonance. In contrast to these three-dimensional systems, two-dimensional confinement of conduction electrons in layered materials featuring distinct electronic structure may lead to unusual Kondo screening at high temperature. In this context, the Fermi level in graphene could easily be tuned relative to the Dirac point to provide the required conduction electrons. Thus, graphene is expected to manifest Kondo screening, and indeed has been observed in defected or transition-metal doped graphene. [11–14] In contrast, the two-dimensional semiconductors cannot naturally host Kondo state due to the lack of conduction electrons, and thus it would be fascinating to design Kondo screening by tuning chemical potential. This will open up a possibility to investigate many-body Kondo physics in reduced dimension including the quantum phase transition between unscreened and screened impurity moment.

Here, we predict a new class of Kondo system in two-dimensional semiconducting phosphorene, which is intrinsic and robust, that are markedly different from the artificial Kondo systems.[4–10] Since the recent exfoliation of single and few-layer phosphorene, it has attracted enormous attention due to anisotropic electronic properties, [15–17] and investigated in the context of many-body physics apart from a plethora of plausible technological applications. [16, 17] A topologically insulating and Dirac semimetal states have been predicted or realized in few-layer phosphorene under external electric field or adatom adsorption. [18–20]

We engineer two-dimensional semiconducting phosphorene such that it becomes susceptible to many-body Kondo screening. While Sc and Cu impurities absorbed on single-layer phosphorene (SLP) triggers semiconductor to metal transition without impurity moment formation, [21, 22] the V, Mn, and Fe impurities generate unscreened moments as the systems remain semiconducting. [21] In sharp contrast, while absorbed at the thermodynamically most stable divacancy defect, Cr impurity retains a localized moment, and also alters the electronic structure of phosphorene such that it becomes susceptible to Kondo screening. Here, we particularly investigate the non-trivial questions that are essential for the appearance of Kondo effect – Impurity absorption site and the concurrent electronic phase transition, impurity spin state, and the interaction between the moment generating impurity state with the host conduction electrons. Strong interaction between the impurity and conduction electrons, which are delocalized along the armchair direction, results into high temperature Kondo state. The impurity moment is quenched at different temperatures in the range 40–200 K. Further, the proposed Kondo state is predicted to be robust against layer thickness and moderate strain, which may be induced by the substrate. We also provide an insight about the Kondo cloud, which has remained experimentally elusive. This is largely due to the fact that the extension of Kondo cloud in conventional metal-impurity system is $\sim 1$ $\mu$m and is much larger than the dimension of current quantum devices. [23–25] In contrast, the Kondo cloud extension in the present system is predicted to be orders of magnitude smaller and could possibly be detected.

## II. CALCULATION DETAILS

Calculations are based on spin-polarized density functional theory (DFT) within the projector augmented wave formalism with an energy cutoff of 500 eV. [26–28] The exchange-correlation energy is described by Perdew-Burke-Ernzerhof functional (PBE). To estimate the robustness of the first-principles predictions, selected calculations are repeated with Heyd-Scuseria-Ernzerhof hybrid functional (HSE06).[29] Hubbard like on-site Coulomb interaction was considered within rotationally invariant approach (DFT+$U_{\rm Cr}$).[30] Single layer and bilayer phosphorene were modelled with 7(zigzag)×5(armchair) and 6(zigzag)×4(armchair) supercells, respectively. A vacuum layer of 15 Å are used, and the structures were optimized until all the force components are below 0.01 eV/Å threshold. Reciprocal space integration was carried out using 4×4×1 Monkhorst-Pack **k**-point grid. We addressed the interlayer interaction in bilayer phosphorene (BLP) with nonlocal van der Waals functional (optB88-vdW) of Langreth and Lundqvist. [31]

## III. RESULTS AND DISCUSSION

The pristine SLP is a semiconductor, and the intrinsic point defects are electronically inactive that does not perturb the transport gap. [21, 32, 33] The divacancy (DV) with 5-8-5 ring structure is the thermodynamically most stable point defect in SLP. While the conventional PBE functional underestimates the DV-SLP gap to 0.97 eV, the hybrid HSE06 functional (1.77 eV) improves the result. In contrast, Cr impurity absorbed at the DV triggers an electronic phase transition, and Cr@DV-SLP becomes half-metallic. The Cr impurity is absorbed in-between the half-layers at the valley site [Figure 1(a) and Appendix A] and form strong covalent bonds through hybridization between P-$p$ and Cr-$d_{x^2-y^2}/d_{yz}$ orbitals. The half-filled electronic shell results in most common $Cr^{2+}/Cr^{3+}$ oxidation states, and the $Cr^{3+}$ ($S$ = 3/2) state is found to be most stable within conventional PBE exchange-correlation functional. The competing $Cr^{2+}$ ($S$ = 2) state is 20 meV higher in energy [Figure 1(b)]. However to proceed further, it is necessary to confirm the intrinsic nature of such electronic phase transition and rule out any artefact arising from the conventional exchange-correlation description.

We have recalculated using the on-site Coulomb interaction $U_{\rm Cr}$, and expensive hybrid HSE06 exchange-correlation functional, which improves the description of $p-d$ interaction. Indeed, the electronic half-metallic phase along with Cr impurity moment are found to be robust within the adopted theoretical hierarchy [Figure 2(a) and Appendix A]. With increasing $U_{\rm Cr}$ (0−4 eV), the $p-d$ hybridization decreases, and is reflected through increasing $\langle \mathrm{P-Cr} \rangle$ distances. A careful study of the partial density of states suggests that the bonding is derived through $p-d_{yz}$ hybridization, which dictates $S$=2 spin state localized at the impurity [Figure 2(b)]. This state is 10, 35 and 10 meV lower in energy than the competing $S$ = 3/2 state, for $U_{\rm Cr}$=2, 4 eV, and HSE06, respectively [Figure 1(b)]. Regardless the impurity spin state, the Cr@DV-SLP is found to be half-metallic, where the gap is closed in the minority spin channel [Figure 2(a) and Appendix A]. The unreconstructed DV in SLP generates four dangling bonds, which are incompletely filled by Cr absorption and produces $Cr^{2+}/Cr^{3+}$ solutions. Therefore, defected SLP becomes hole-doped and consequently the Fermi level shifts lower in energy resulting in gap closing. The partial density of states corroborates this picture, where the $E_F$ lies within the band that is arising from the P-$p$ electrons [Figure 2(a)]. Further, the charge density corresponding to the conduction electrons near the $E_F$ indicates that these conduction electrons are delocalized over the entire supercell (∼ 23 Å), along the armchair direction [Figure 2(c)] with very high carrier mobility. We note that the electronic phase transition is of Lifshitz type, [34] and a hole-pocket emerges at the Γ-point [Figure 2(a)]. Such Lifshitz transition triggers interesting electronic responses. [35–37] Although Lifshitz transition has been reported for single and few-layer phosphorene, [19, 20, 35, 36] in majority of these cases the Fermi surface topology changes due to neck formation. In the present case for Cr@DV-SLP, the Lifshitz transition arises due to pocket formation, and precedes Kondo screening.

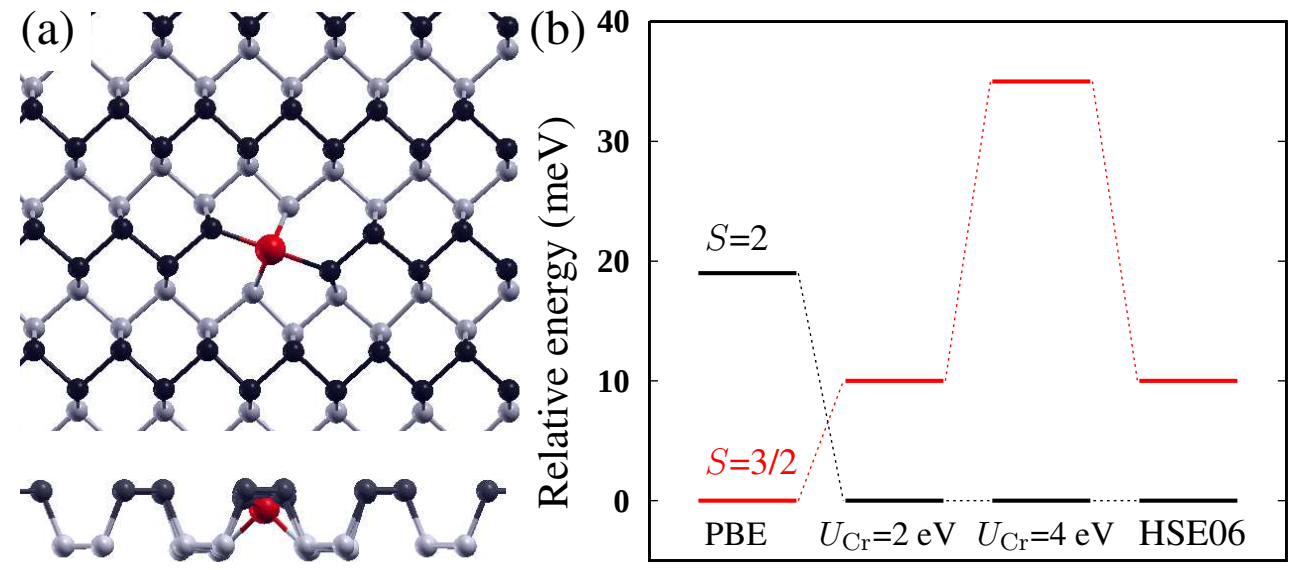

FIG. 1. (a) Top and side view of Cr absorption at the DV(5|8|5) on SLP. The Cr impurity (red ball) is absorbed in-between the two half-layers indicated in black and white balls, at the valley site (Appendix A). (b) The Cr impurity is found to have competing spin-3/2 and spin-2 states. White the conventional PBE exchange-correlation predicts a spin-3/2 ground state, inclusion of on-site Coulomb interaction and hybrid exchange-correlation anticipate the spin-2 state as the ground state.

The corresponding impurity moment is localized on the absorbed Cr, as it is evident from the magnetization density [Figure 2(b)]. The high-spin $S$=2 ground state, within supplemented electron correlation and hybrid exchange-correlation functional approaches [Figure 1(b)], arises from half-filled $d_{z^2}$, $d_{xy}$, $d_{x^2-y^2}$ and $d_{xz}$ orbitals, which are non-degenerate due to the lack of symmetry around Cr-impurity. In contrast, the competing $S = 3/2$ impurity state arises from the $d_{z^2}$, $d_{xy}$, and $d_{xz}$ orbitals. The presence of such localized Cr moment along with P-$p$ conduction electrons indicates the possibility of Kondo state (Figure 2). However, for the occurrence of Kondo resonance, the magnetic impurity levels must have sizeable electronic coupling with the host conduction electrons, and must not behave as free moment. In this context, we have calculated the partial crystal orbital Hamilton population, [38] which indicate substantial hybridization between the moment generating Cr-$d$ orbitals and host conduction P-$p$ electrons (Appendix A). Further, the hybridization $\Delta_\nu$ is quantified from the partial density of states (pDOS) $\rho_\nu$ using a Kramers-Kronig relation, [39]

$$\Delta_\nu(\epsilon) = -\mathrm{Im}\left[\int d\epsilon' \frac{\rho_\nu(\epsilon')}{\epsilon - \epsilon' - i0}\right]^{-1}.$$

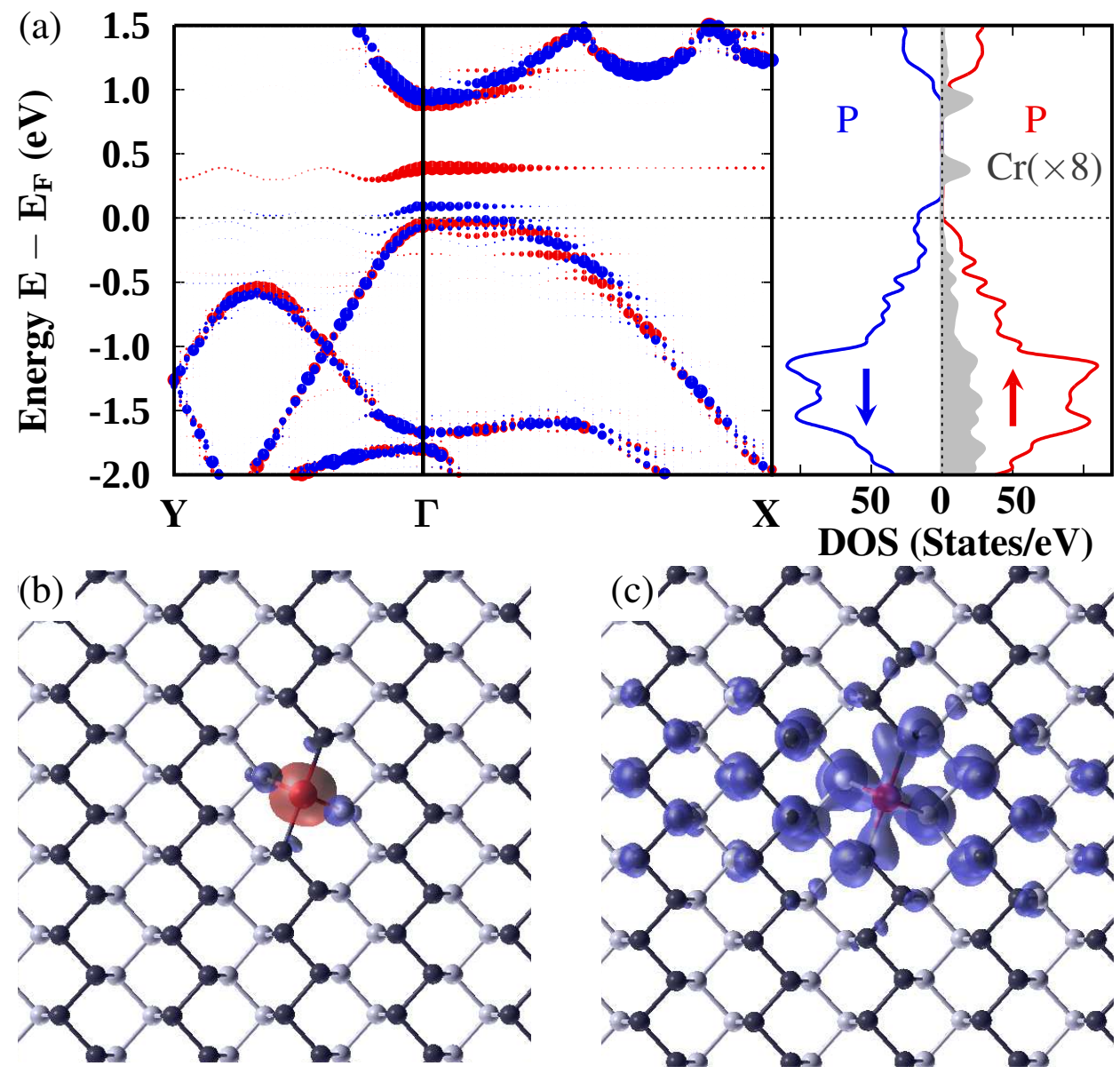

FIG. 2. (a) Unfolded band structure and the corresponding density of states (DOS) of Cr@DV-SLP with very small ∼0.7% Cr-impurity concentration, calculated with DFT+$U_{\mathrm{Cr}}$ formalism with $U_{\mathrm{Cr}}$=4 eV. Red (blue) color represents the majority (minority) channel. While the anisotropic dispersion along the armchair and zigzag directions is retained, the defected Cr@DV-SLP becomes half-metallic. Further, the conduction electrons are of P-$p$ character. (b) The magnetization density indicates that the Cr magnetic moment is mainly localized at the impurity site. (c) The corresponding charge density indicates that the conduction electrons are delocalized along the armchair direction and over the entire supercell (∼ 23 Å).

Calculated $\Delta_\nu(\epsilon)$ shows strong interaction between the localized $d$-orbital with the conduction electron as shown in Appendix A and will be useful to estimate $T_K$.

Thus, the Cr@DV-SLP system has all the ingredients required to form Kondo state at low temperature: local Cr impurity moment, host conduction electrons, and a sizeable coupling between them. However, the nature of Kondo state and a correct model are very difficult to predict and non-trivial. For the classic case of Fe impurities in gold and silver, the correct model description has only been determined very recently, [40] which was first manifested in 1930s. [1, 3] Although, the spin-orbit coupling is not considered in the present calculations, it may be expected to be negligible as in the case of Co-absorbed graphene. [11] Such scenario is expected to generate conventional spin-only SU(2) Kondo state. Further, as all the half-filled impurity $d$-states interact strongly with the conduction electrons, the impurity moment will be completely screened, and a four-channel spin-2 Kondo state is expected. However, the competing spin-3/2 impurity state is only 10-30 meV higher in energy depending on theoretical hierarchy. Thus, one may not completely rule out a spin-3/2 three-channel Kondo state. Further, we estimate the Kondo temperature $T_K$ for the screening of each half-filled Cr-$d$ orbitals corresponding to the spin-2 state ($U_{\mathrm{Cr}}$=4 eV) by using, [41]

$$k_B T_K \simeq \sqrt{2\Delta_\nu \frac{\mathcal{U}_\nu}{\pi}} \exp\left[-\frac{\pi}{2\Delta_\nu}\left(\frac{1}{|\epsilon_\nu|} + \frac{1}{|\epsilon_\nu + \mathcal{U}_\nu|}\right)^{-1}\right],$$

where $\epsilon_\nu$ is the energy of the $d$-level (∼ 2.2 eV), $\mathcal{U}_\nu$ is the Coulomb interaction energy that are calculated from pDOS $\rho_\nu$, and $\Delta_\nu$ ranges between 0.10–0.30 eV (Appendix A). The estimated $T_K$ is found to be high between 40–200 K, where the non-degenerate $d$-orbitals form Kondo singlet at different temperatures, and the impurity moment is completely quenched at multi-stage. We anticipate ellipsoidal Kondo cloud elongated along the armchair direction, and estimate the coherence length $\xi_K = \hbar v_F / k_B T_K \sim$ 20–4 nm, where Fermi velocity $v_F \sim 10^5$ m/s and $T_K \sim$ 40–200 K, which is orders of magnitude smaller than a typical $\xi_K \sim 1$ $\mu$m in conventional metal-impurity systems. Thus, it could be finally possible to experimentally detect the Kondo cloud in the present system, which remained elusive till today. [25, 42, 43]

Having discussed the strategy to design intrinsic Kondo state in free-standing semiconducting SLP, we next discuss its robustness on a substrate and with increasing layer thickness. The substrate effect is incorporated via induced strain in SLP, while the plausible effect of substrate on the intrinsic electronic structure of SLP has not been considered, and is out of the scope of this paper. The effect of increasing thickness has been investigated through bilayer phosphorene (BLP).

First, we discuss the robustness of Kondo state in strained SLP, as an experimental design of Kondo state could be hindered by the substrate induced strain. Although, pristine SLP is known to withstand an uniaxial tensile strain up to 30% due to its puckered structure, the electronic structure is found to be amenable. [44–46] Further, ripples and corrugations in large area SLP could lead to strained regions. [47] Thus, it is prudent to investigate the robustness of Kondo state in strained SLP. We have considered a moderate $\varepsilon_{a/z} = \pm$ 4% uniaxial strain along both armchair and zigzag directions. We find that straining the SLP along the armchair direction is easier than the zigzag direction (Appendix B). Such behavior is anticipated as the P−P bonds along the armchair direction are weaker, and thus it is energetically less expensive to distort the structure along this direction. The strain relaxation in the transverse direction is found to be complex. [48] While the lattice relaxation along the transverse zigzag direction is found to be negligible under applied $\varepsilon_a$, the distance between the half-layers is affected significantly. In contrast, for applied $\varepsilon_z$ the lattice relaxation along armchair direction is proportionate, while the distance between the two half-layers remains unperturbed. Therefore, strain in the lattice is expected to alter the intrinsic electronic structure of SLP. Generally for pristine SLP with $-4\% \leqslant \varepsilon_{a/z} \leqslant 4\%$, the gap increases with uniaxial tensile strain, while a compressive strain reduces it (Appendix B). The gap is calculated to be 0.75 to 1.05 eV within the applied strain range, and the results are in agreement with previous reports. [44] Further, the DV-SLP remains semiconducting under applied strain with 0.8–1.0 eV gap (Appendix B).

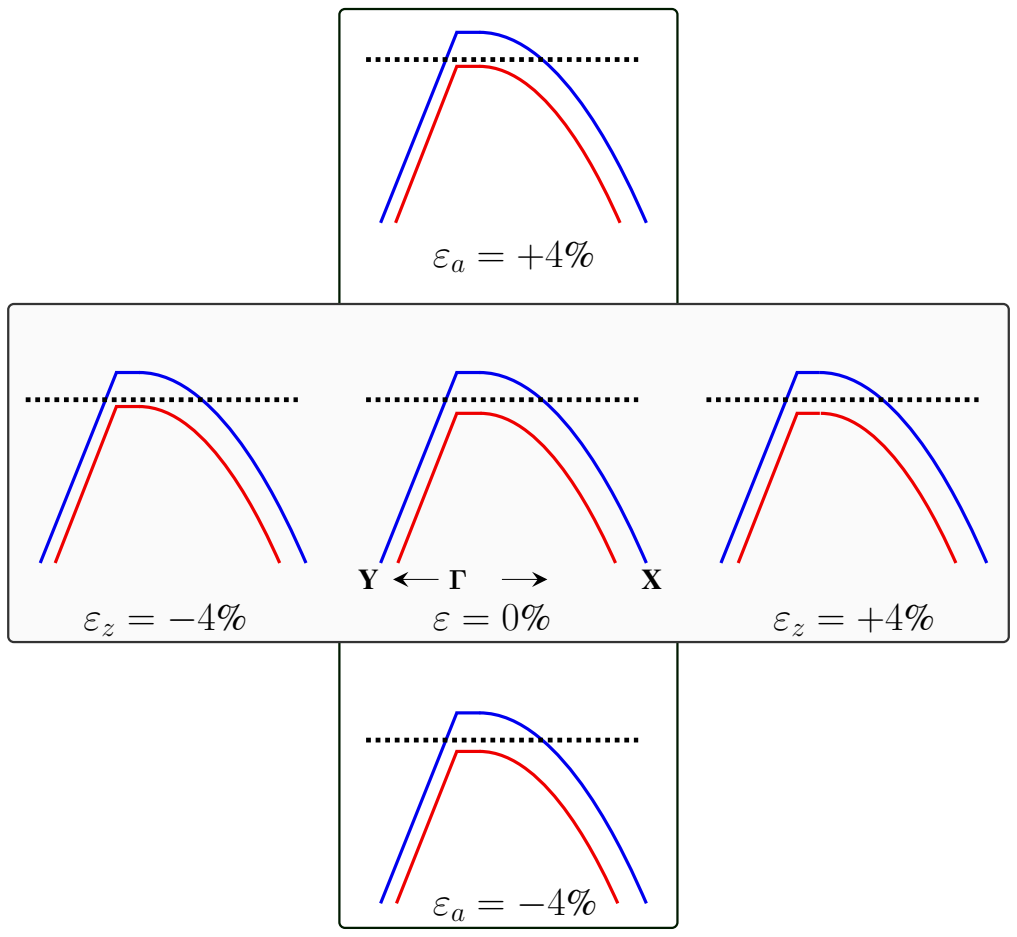

FIG. 3. Schematic band structure of Cr@DV-SLP under moderate uniaxial tensile and compressive strains along armchair $\varepsilon_a = \pm 4\%$ and zigzag $\varepsilon_z = \pm 4\%$ directions. Similar to the unstrained Cr@DV-SLP, the electronic structure remains half-metallic under uniaxial strains, and the conduction electrons are delocalized along the armchair directions. Further, the conventional PBE functional and supplemented on-site Coulomb interaction of $U_{\rm Cr} =$ 2 and 4 eV indicate qualitatively similar trend.

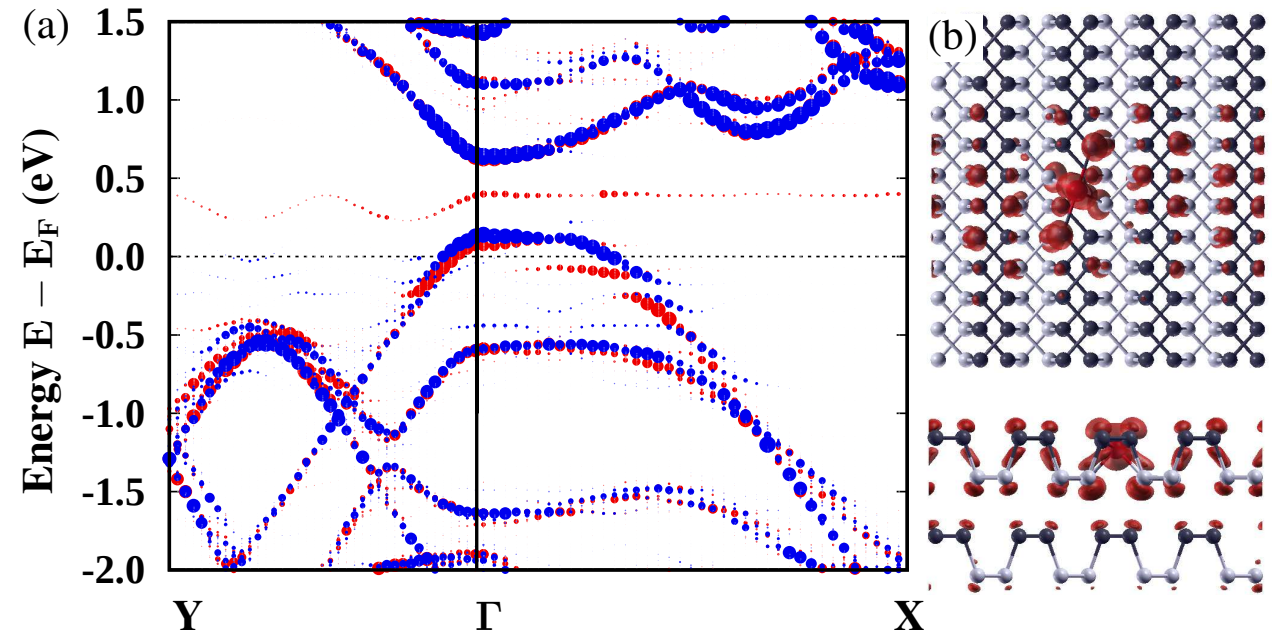

FIG. 4. (a) While the defected DV-BLP is semiconducting, the Cr-impurity absorption at DV transforms the electronic structure to metallic. (b) The P-$p$ conduction electrons are delocalized along the armchair direction in the defected layer, and over the entire supercell considered. Although the results are shown for $U_{\rm Cr}$=4 eV, the PBE and $U_{\rm Cr}$=2 eV provide qualitatively similar results.

Further, the magnetic and electronic structures of strained Cr@DV-SLP are strikingly similar with the unstrained counterpart that are discussed earlier. The overall magnetic structure of absorbed Cr impurity remains unaffected in the strained DV-SLP. The comparative stability of competing spin-3/2 and spin-2 impurity states is strikingly similar, as it has been discussed for unstrained SLP. For the entire $-4\% \leqslant \varepsilon_{a/z} \leqslant 4\%$ range, the spin-2 impurity state is found to be the ground state with $U_{\rm Cr} \geqslant$ 2 eV, where the $d_{z^2}$, $d_{xy}$, $d_{xz}$ and $d_{x^2-y^2}$ orbitals generate the impurity moment. The Cr@DV-SLP remains half-metallic under the uniaxial strain (Figure 3), and the conduction electrons are delocalized along the armchair direction. Further, the absorption site for Cr remains unchanged in strained DV-SLP, along with a large Cr absorption energy (-3.8 to -3.5 eV). The COHP and $\Delta_\nu(\epsilon)$ calculations indicate a sizeable coupling between the moment generating Cr-$d$ orbitals with the host P-$p$ conduction electrons. Thus, Cr@DV-SLP under moderate strain survives with all the features required for the appearance of Kondo state.

The experimental fabrication of SLP might be a challenging task compared to a few-layer phosphorene. Thus, we investigate the effect of layer thickness on the Kondo state. In this regard, we have investigated BLP with energetically preferable AB stacking, which is a natural order in black phosphorus. Calculated lattice parameters, 3.33 and 4.52 Å along the zigzag and armchair directions, respectively; and the interlayer separation of 3.20 Å are in good agreement with the previous results. [49, 50] The band gap decreases in pristine BLP (0.55 eV within the PBE exchange-correlation functional), in agreement with earlier results. [19, 49, 51] Similar to SLP, the DV defect is also electronically inactive in BLP with 0.57 eV gap. The Cr impurity is absorbed at the valley site, with strong binding energy of −3.10 eV ($S$=2 with $U_{\rm Cr}$=4 eV). Moreover, the intrinsic electronic structure of DV-BLP is completely altered and the Cr@DV-BLP becomes metallic [Figure 4(a) and Appendix C], with delocalized conduction electrons along the armchair direction [Figure 4(b)]. Further, the absorbed Cr generates a localized impurity moment. The energy difference between the competing $S$=3/2 and $S$=2 spin states is found to be -44, -21 and 2 meV for $U_{\rm Cr}$ = 0, 2, and 4 eV, respectively. For both the spin solutions the moment generating Cr-$d$ states have strong hybridization with the conduction electrons. In this case, the non-degenerate $d$-moments are quenched at temperatures in the range 18 - 80 K. Therefore, we infer that increasing layer thickness will not hinder the possibility of Kondo state in bilayer and few-layer phosphorene.

## IV. CONCLUSIONS

In summary, within the first-principles calculations, we rigorously report the evidences of intrinsic, robust and high-temperature Kondo screening in Cr absorbed at a thermodynamically stable defect site in semiconducting phosphorene. The semiconductor to (half) metal phase transition is induced by Cr impurity. In addition, a competing spin-3/2 or spin-2 localized state is generated at the Cr impurity. Further, the conduction electrons are delocalized along the armchair direction, and strongly interact with the moment generating Cr-$d$ orbitals at the impurity site. We infer these picture leads to a spin-only SU(2) Kondo state. A multi-channel screening is predicted, and the moments at the non-degenerate $d$-orbitals are quenched at different temperatures in the range 40 – 200 K. Further we show that such Kondo state is robust under layer thickness, and moderate strain, which could be induced by substrate. Such robustness will simplify its experimental realization. While the present investigation addresses the magnetic screening with vanishingly small impurity concentration, it would be interesting to study the competition and crossover between screening and RKKY-induced magnetic order with increasing impurity concentration. We hope the present results will trigger further theoretical and experimental studies.

## ACKNOWLEDGMENTS

We thank G. J. Sreejith for useful discussions, and acknowledge the supercomputing facilities at the Centre for Development of Advanced Computing, Pune; Inter University Accelerator Centre, Delhi; and at the Center for Computational Materials Science, Institute of Materials Research, Tohoku University. M. K. acknowledges the funding from the Department of Science and Technology, Government of India under Ramanujan Fellowship, and Nano Mission project SR/NM/TP-13/2016.

## Appendix A: Cr@DV-SLP

A divacancy (DV) in puckered phosphorene lattice can be created by removing two adjacent atoms from the same or different half-layers. The DV of the first kind is known to be thermodynamically most stable with DV(5|8|5) configuration with 1.33 eV formation energy. While there are many other possible DV configurations, their formation energies are found to be higher, which has been discussed in literature in great details. [21, 33, 52, 53] A different divacancy configuration (5|7|5|7) was reported, [54] however, we find this structure to be 113 meV higher in energy.

At the most thermodynamically stable DV(5|8|5) configuration, the Cr impurity is absorbed between the two half-layers such that Cr binds with P atoms from both defected and un-defected half-layers. We report two competing spin states $S$ = 3/2 ($Cr^{3+}$) and $S$ = 2 ($Cr^{2+}$)

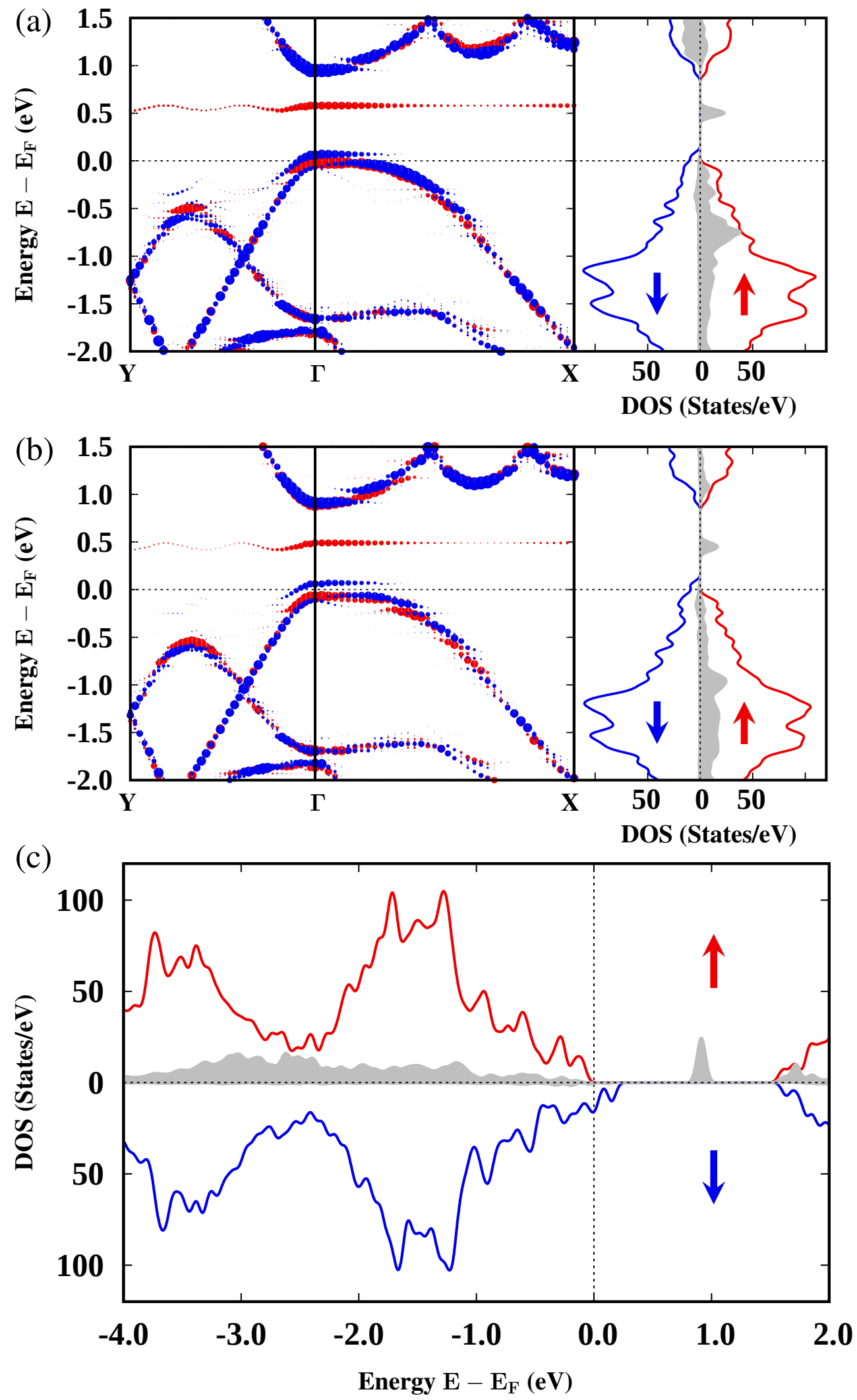

FIG. 5. Unfolded band structure and the corresponding density of states of Cr@DV-SLP calculated with (a) PBE exchange-correlation functional, (b) On-site Coulomb interaction $U_{\rm Cr}$ = 2 eV, and (c) Density of states of Cr@DV-SLP calculated with HSE06 hybrid exchange-correlation functional. All calculations are performed for $S$ = 2 spin ground state. The half-metallic character is consistent within all the considered theoretical hierarchy.

TABLE I. Structural details of Cr absorption, which is influenced by supplemented Coulomb interaction for Cr impurity, $U_{\mathrm{Cr}}$ for their respective ground state configurations. The depth is measured from the defected half-layer. In response to an increasing Coulomb interaction, the Cr–P bonds are elongated.

| Coulomb interaction | Cr–P (Å) bonds | Depth (Å) |
|---|---|---|
| PBE | 2.38–2.44 | 0.66 |
| $U_{\mathrm{Cr}} = 2$ eV | 2.47–2.48 | 0.58 |
| $U_{\mathrm{Cr}} = 4$ eV | 2.51–2.54 | 0.52 |

within different theoretical hierarchy. For finite Coulomb interaction at the Cr-impurity, the later was found to be the ground state, which is also the case for expensive hybrid HSE06 functional. This has been discussed in the main text [Figure 1(b)]. The preferred oxidation state of Cr depends on the Cr–P distances. The spin-2 solution is preferred for finite Coulomb interaction due to increasing $\langle \mathrm{Cr} - \mathrm{P} \rangle$ distances. Subsequently, Cr moves closer to the defected half-layer (Table I).

It is known that point defects do not alter the intrinsic semiconducting nature of phosphorene. [21, 33] In contrast, Cr absorption at the DV induces electronic phase transition, and becomes half-metallic, as shown in Figure 5(a) within PBE exchange correlation functional. The picture remains same while explicit on-site Coulomb interaction is added to the PBE functional or within hybrid HSE06 calculations as shown in Figures 5(b) and (c), respectively. This indicates the semiconductor to half-metal transition is robust and not an artefact of underlying theoretical approximations. Similar to the $S = 2$ ground state as discussed above, the electronic structure of the competing $S = 3/2$ spin state is also found to be half-metallic within all theoretical descriptions of exchange-correlation functional employed here, and shown for $U_{\mathrm{Cr}} = 4eV$ in Figure 6. Note that the $S = 3/2$ spin state lies only 10-35 meV higher in energy within these approaches.

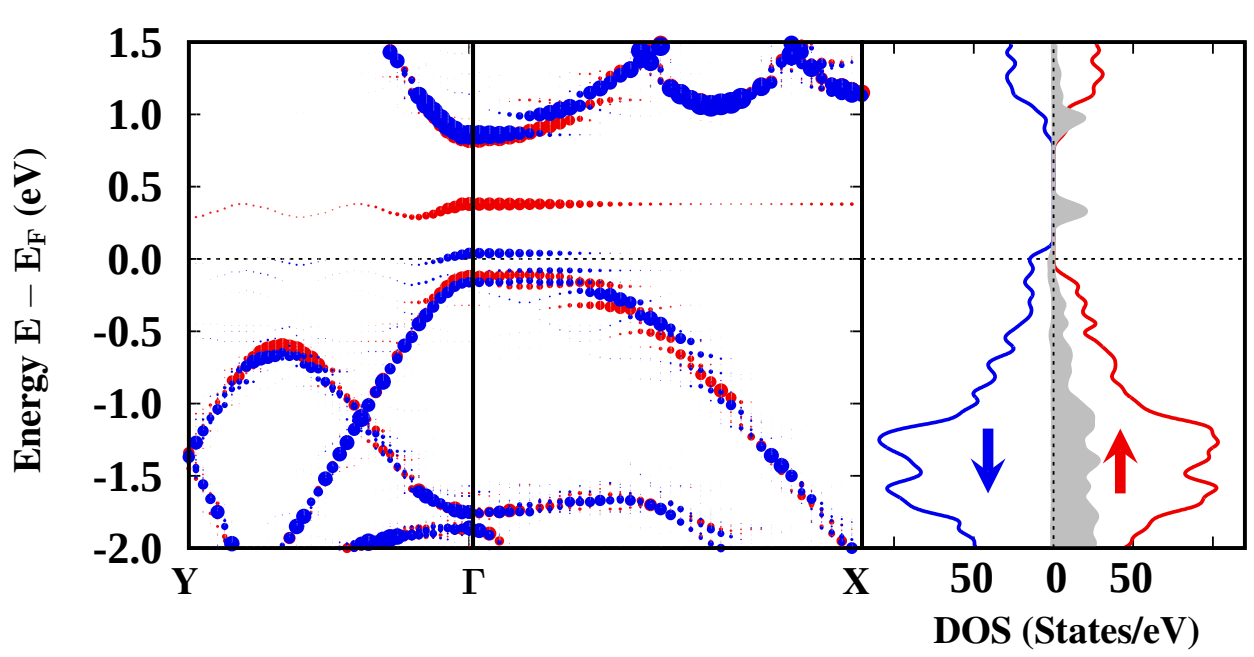

FIG. 6. Unfolded band structure and the corresponding density of states of Cr@DV-SLP for the low-lying $S = 3/2$ state, calculated for an on-site Coulomb interaction $U_{\mathrm{Cr}} = 4$ eV. The Cr@DV-SLP remains a half-metallic regardless the spin state.

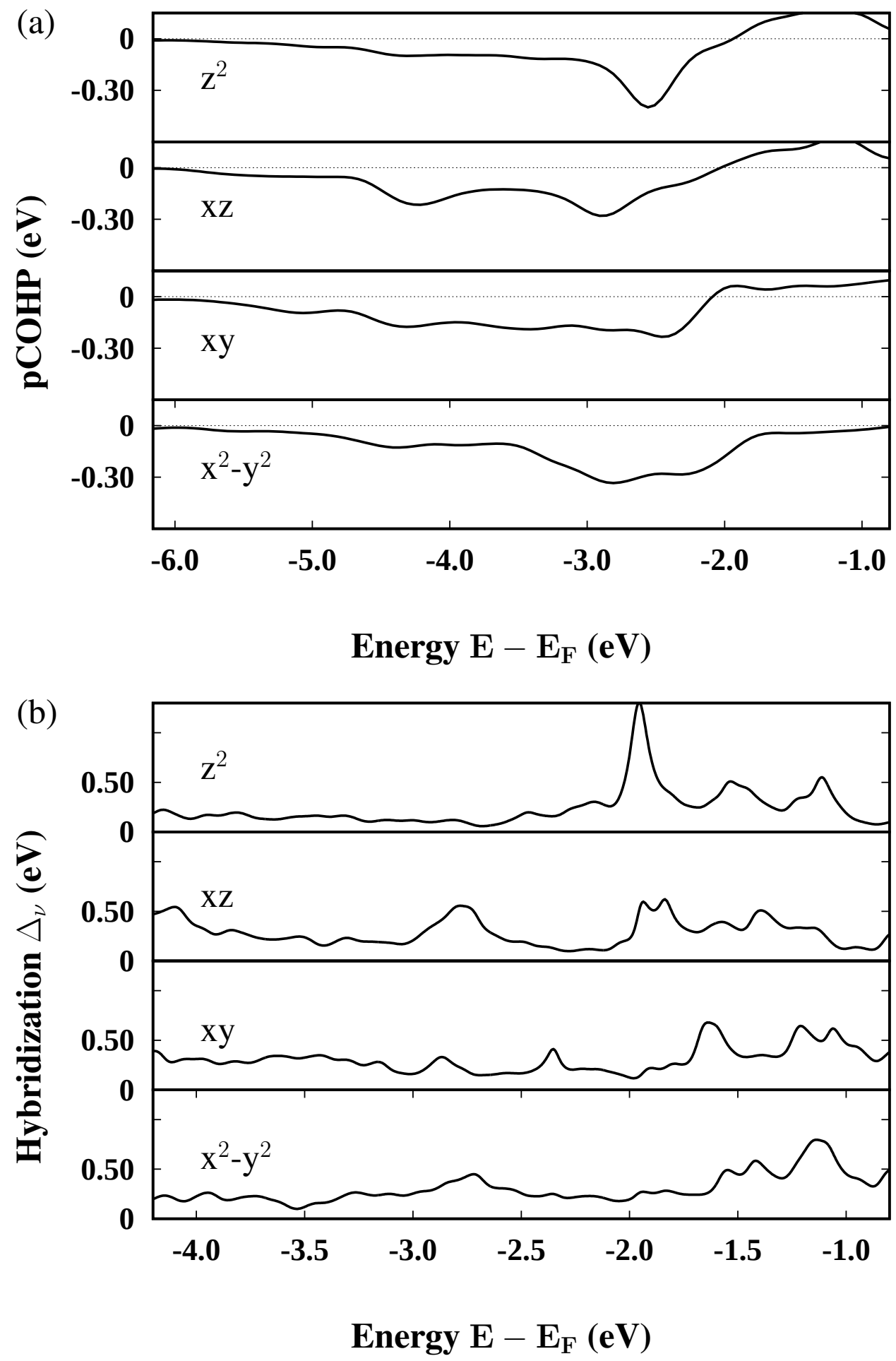

FIG. 7. (a) Projected crystal orbital Hamilton population (pCOHP) between Cr-$d$ and P-$p$ electrons is evaluated for $U_{\mathrm{Cr}} = 4$ eV. The negative (positive) pCOHP values refer to bonding (anti-bonding) nature of interaction between Cr-$d$ and P-$p$ electrons. The presence of bonding interaction for all magnetic $d$ orbitals with the P-$p$ electrons favors screening of all moments during Kondo state formation. (b) Hybridization $\Delta_\nu$ evaluated using the Kramers-Kronig relation for Cr-$d$ orbitals in case of $U_{\mathrm{Cr}} = 4$ eV.

Note that the electronic band structures for the supercells are calculated using the band unfolding technique implemented in the BandUP code [55, 56]. For a supercell, the band structure calculation results in zone folding of bands due to the contraction of corresponding Brillouin zone, which makes the identification of impurity states from the host bands difficult. Here, the band unfolding is carried out by mapping the spectral weights of supercell Brillouin zone onto the primitive cell Brillouin zone.

In addition to the localized moment at the Cr impurity, it is necessary to have a strong interaction between this localized moment with the conduction electrons from the host phosphorene for the manifestation of many-body Kondo state. We consider the crystal orbital Hamilton population (COHP) scheme for identifying the bonding interaction between Cr-$d$ and P-$p$ states. [38, 57] Here, the wave function calculated using plane-wave method is utilized to evaluate the density and Hamiltonian matrix elements, which are then projected over local orbitals. The resultant picture can distinguish between bonding, non-bonding, and anti-bonding interactions between atoms. In case of Cr@DV-SLP, we observe bonding states in the vicinity of the band centers of each $d$ orbital contributing to the impurity moment. The presence of such hybridized states supports strong interaction between Cr-$d$ and P-$p$ states [Figure 7 (a)].

TABLE II. Band centres $\epsilon_\nu$ and Coulomb repulsion energies $\mathcal{U}_\nu$ are calculated from the partial density of states for $S = 2$ spin ground state state with $U_{\rm Cr} = 4$ eV. The hybridization $\Delta_\nu$ is evaluated using the Kramers-Kronig relation described in the main text. The orbital Kondo temperatures are calculated using these parameters. The non-degenerate $d$-orbitals at Cr impurity are quenched at different temperatures.

| Orbital $\nu$ | $\nu$–band center $\epsilon_\nu$ (eV) | Coulomb repulsion $\mathcal{U}_\nu$ (eV) | Hybridization $\Delta_\nu$ (eV) | $T_K$ (K) |
|---|---|---|---|---|
| $z^2$ | -2.21 | 3.15 | -0.276 | 202 |
| $xz$ | -2.20 | 2.53 | -0.120 | 119 |
| $xy$ | -2.27 | 3.22 | -0.204 | 43 |
| $x^2 - y^2$ | -2.16 | 3.12 | -0.227 | 78 |

Further, the hybridization between the unpaired Cr-$d$ orbitals with the P-$p$ conduction electrons are calculated using the Kramers–Kronig relation (described in the main text) from the projected density of states generated from the plane-wave calculations [Figure 7 (b)]. We estimate the temperature $T_K$ below which Kondo resonance is anticipated for the concerned $d$ orbital. Here, we calculate the relevant quantities from the projected density of states (Table II).

### Appendix B: Cr@DV-SLP under strain

To check the robustness of Kondo state, we have investigated the Cr@DV system under strain, which many be induced while SLP is placed on a substrate. In this regard, we first discuss the strained SLP under moderate uniaxial strains $\varepsilon_{a/z} = \pm\, 4\%$ along the armchair and zigzag directions. We have calculated the strain energy while relaxing the lattice in the transverse direction, and the corresponding strain energy is calculated as,

$$E_S = \frac{\mathscr{E}_{\rm strained} - \mathscr{E}_{\rm pristine}}{A_{\rm pristine}},$$

where $\mathscr{E}_{\rm strained}$ and $\mathscr{E}_{\rm pristine}$ are the total energy of uniaxially strained and pristine supercell, while $A_{\rm pristine}$ is the cross-section of pristine supercell. Calculated strain energy is shown in Figure 8. We find that applied strain could easily be accommodated along the armchair direction due to puckering, than the zigzag direction. This is reflected from the higher strain energy along the zigzag direction compared to the same along the armchair direction. This is attributed to weaker bonding between P atoms along this direction.

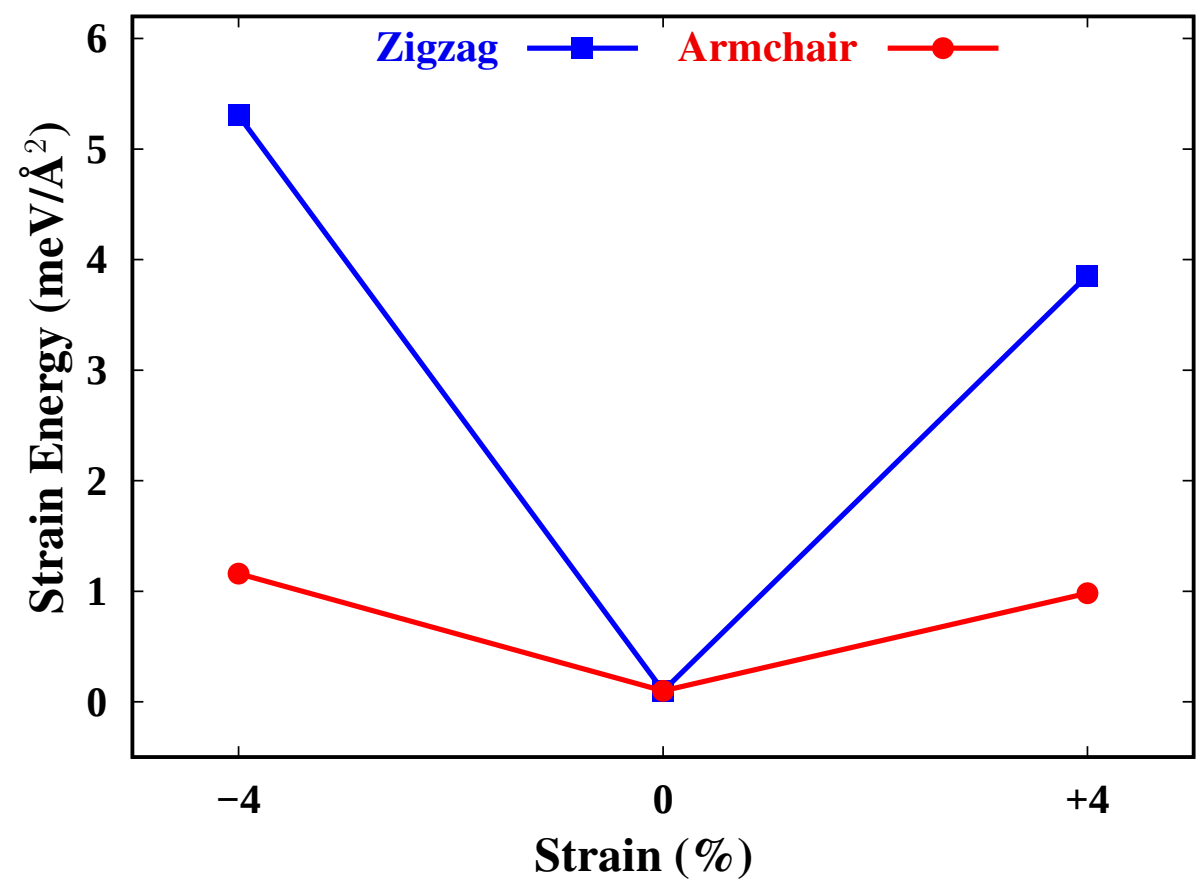

FIG. 8. Strain energy of pristine SLP under uniaxial strain applied along armchair and zigzag directions. Straining the phosphorene lattice along the armchair direction is easier compared to the zigzag direction.

The calculated band gaps are modified in the presence of uniaxial strain for both pristine and defected DV-SLP (Figure 9). We observe that the compressive (tensile) strain leads to decrease (increase) in band gap for pristine SLP. Further, the band gaps are found to be direct in nature for all cases except for $\varepsilon_z = -4\%$. Most importantly, the intrinsic semiconducting nature remains intact for DV-SLP for considered strain range (Figure 9). Although the DV-SLP behave in qualitatively similar fashion as the pristine SLP, the calculated gap for $\varepsilon_z = 4\%$ is found to be lower than its unstrained counterpart, which is in agreement with the previous report. [58]

It would be interesting to investigate how strain affects the absorption energy of Cr-impurity at the DV. While the Cr absorption energy is found to be $\sim 3.8$ eV for zero strain, applied strain does not significantly affect the binding energy, and remains in the 3.8–3.5 eV range for the $\varepsilon_{a/z} = \pm\, 4\%$ strain range (Figure 10). However, we observe some small feature with increasing strain – In general, the absorption energy decreases with increasing strain except for $\varepsilon_a = 4\%$ which almost remains unaltered.

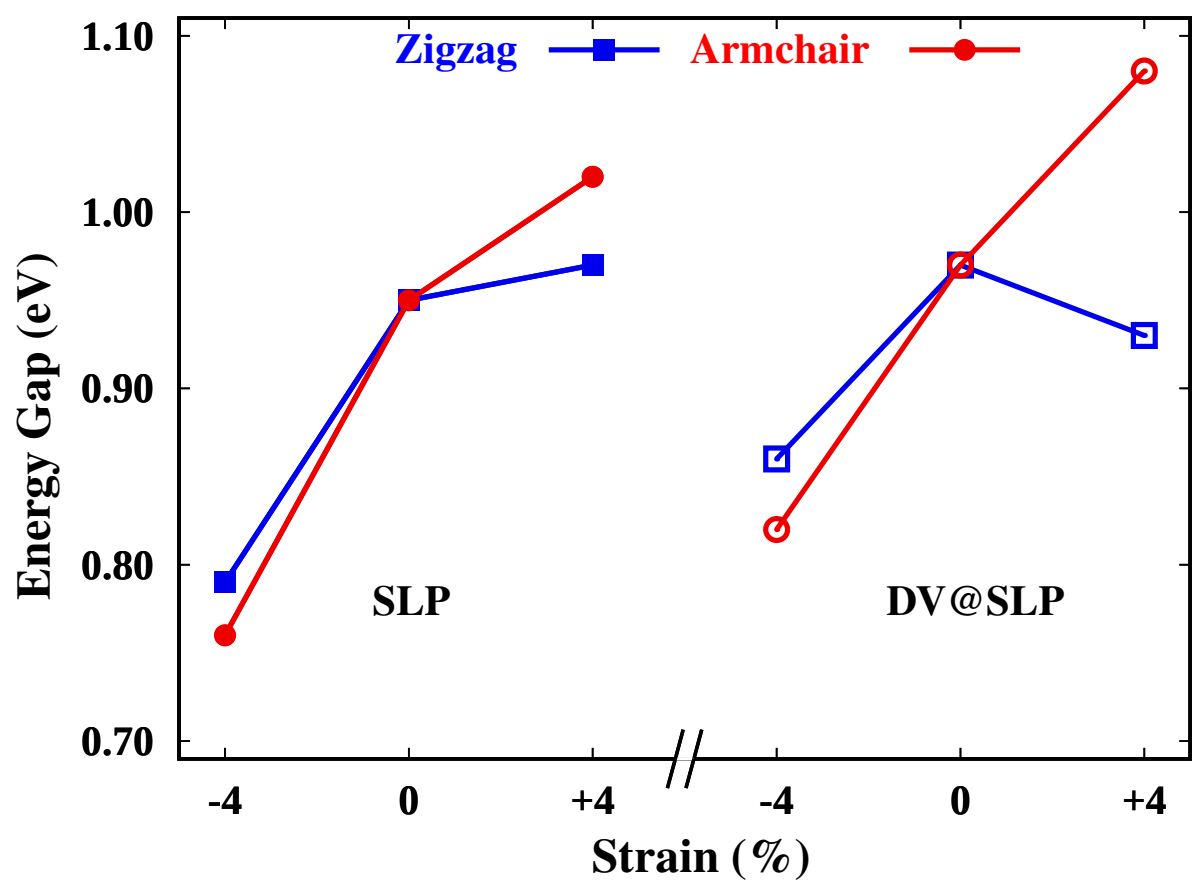

FIG. 9. Calculated energy gaps for pristine SLP (left) and DV-SLP (right) under uniaxial strain applied along armchair and zigzag directions using the conventional PBE functional.

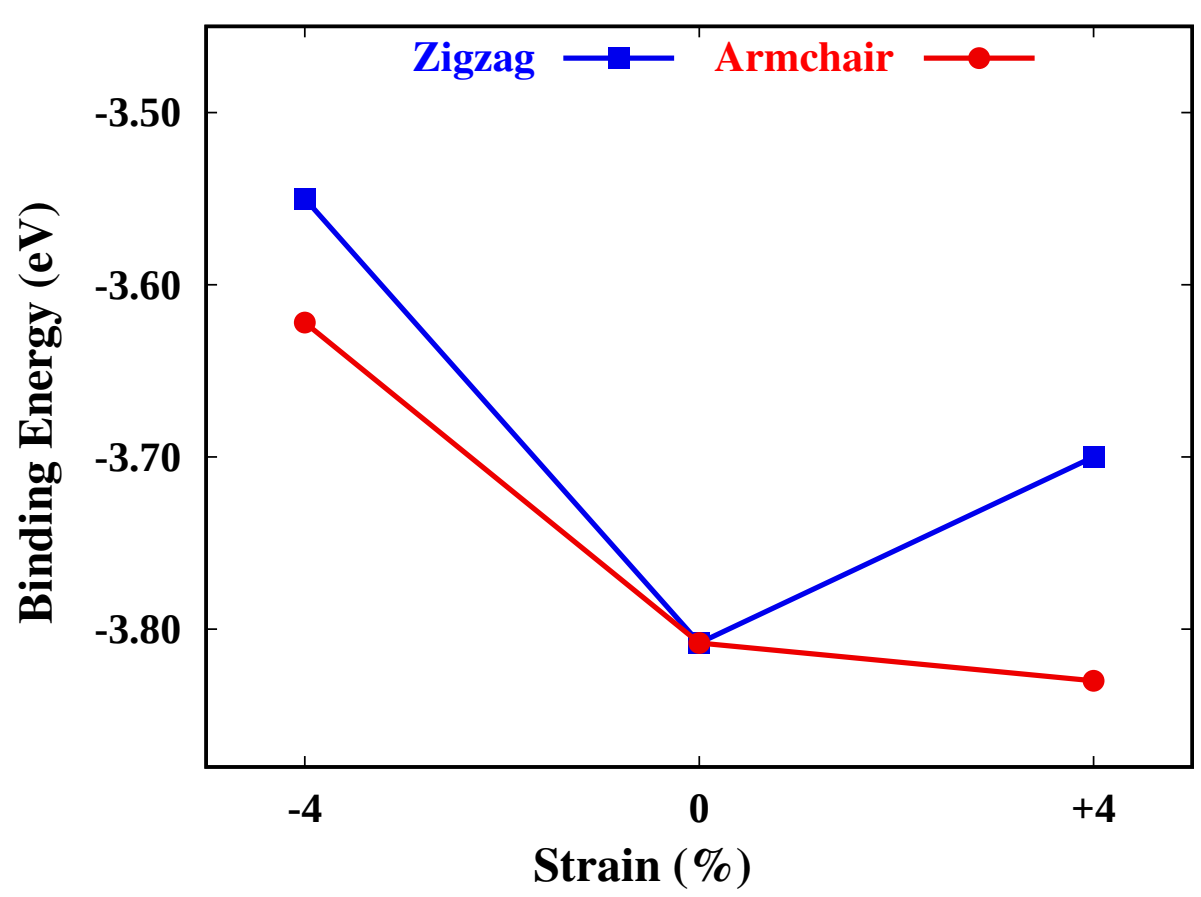

FIG. 10. Binding energy for Cr@DV-SLP. The applied strain does not affect the binding energy significantly, which remains in the 3.80–3.50 eV range.

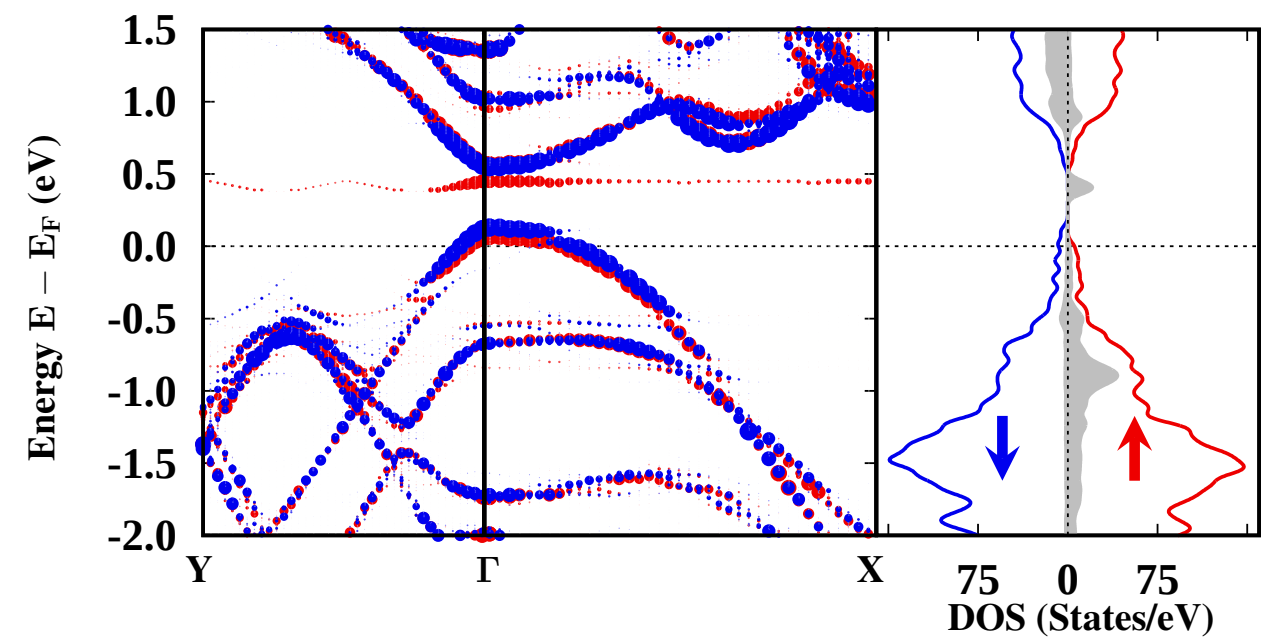

FIG. 11. (a) Unfolded band structure and the corresponding density of states of Cr@DV-BLP calculated with PBE exchange-correlation functional. Unlike the half-metallic nature of Cr@DV-SLP, the Cr impurity absorption in DV-BLP leads to a metallic solution.

### Appendix C: Cr@DV-BLP

To further substantiate the robustness of the Kondo state for few-layer phosphorene, we have investigated the effects of layer thickness through the bilayer phosphorene (BLP). It is already known that the band gap decreases with increasing layer thickness, which might help semiconductor to metal electronic crossover while Cr is absorbed at a DV on BLP. Indeed, regardless the value for the on-site Coulomb interaction, Cr@DV-BLP is found to be metallic. While the calculated band structure for $U_{\mathrm{Cr}}$=4 eV is shown in Figure 4, the Figure 11 shows the same for the conventional PBE exchange-correlation functional.

* Corresponding author: mukul.kabir@iiserpune.ac.in